\documentclass{emulateapj}

\usepackage{amssymb}

\def\ts     {\thinspace}

\def\etal   {{\rm et\ts al.}}

\def\nii   {[NII]$_{\rm 205 \mu m}$}

\slugcomment{Accepted for publication in the Astrophysical Journal Letters}

\shorttitle{Search for \nii\ at z=6.42}
\shortauthors{Walter et al.}

\begin{document}

\title{A sensitive Search for \nii\ Emission in a z=6.4 Quasar Host
  Galaxy}

\author{Fabian Walter\altaffilmark{1}, Axel Wei{\ss}\altaffilmark{2},
Dominik A. Riechers\altaffilmark{3,7}, Christopher
L. Carilli\altaffilmark{4}, Frank Bertoldi\altaffilmark{5}, Pierre
Cox\altaffilmark{6}, Karl M. Menten\altaffilmark{2}}

\altaffiltext{1}{Max-Planck-Institut f\"ur Astronomie, K\"onigstuhl
  17, Heidelberg, D-69117, Germany}

\altaffiltext{2}{Max-Planck-Institut f\"ur Radioastronomie, Auf dem
H\"ugel 69, Bonn, D-53121, Germany}

\altaffiltext{3}{Astronomy Department, California Institute of
Technology, MC 105-24, 1200 East California Boulevard, Pasadena, CA
91125, USA}

\altaffiltext{4}{National Radio Astronomy Observatory, PO Box O,
Socorro, NM 87801, USA}

\altaffiltext{5}{Argelander-Institut f\"ur Astronomie, Universit\"at
Bonn, Auf dem H\"ugel 71, Bonn, D-53121, Germany}

\altaffiltext{6}{Institut de RadioAstronomie Millim\'etrique, 300 Rue
  de la Piscine, Domaine Universitaire, F-38406 Saint Martin
  d'H\`eres, France}

\altaffiltext{7}{Hubble Fellow}

\email{walter@mpia.de}

\begin{abstract}
  We present a sensitive search for the $^3\!P_1\rightarrow^3\!P_0$
  ground state fine structure line at 205 microns of ionized nitrogen
  (\nii) in one of the highest redshift quasars (J1148+5251 at z=6.42)
  using the IRAM 30\,m telescope. The line is not detected at a
  (3$\sigma$) depth of 0.47 Jy\,km\,s$^{-1}$, corresponding to a \nii\
  luminosity limit of L$_{\rm [NII]}<$4.0$\times10^8$\,L$_{\odot}$ and
  a L$_{\rm [NII]}$/L$_{\rm FIR}$ ratio of $<2\times10^{-5}$. In
  parallel, we have observed the CO(J=6--5) line in J1148+5251, which
  is detected at a flux level consistent with earlier interferometric
  observations.  Using our earlier measurements of the [CII]
  158\,micron line strength, we derive an upper limit for the
  \nii/[CII] line luminosity ratio of $\sim$1/10 in J1148+5251.
  Our upper limit for the [CII]/\nii ratio is similar to the value
  found for our Galaxy and M\,82 (the only extragalactic system where
  the \nii\ line has been detected to date).  Given the non--detection
  of the \nii\ line we can only speculate whether or not high--z
  detections are within reach of currently operating observatories.
  However, \nii\ and other fine strucure lines will play a critical
  role in characterizing the interstellar medium at the highest
  redshifts (z$>$7) using the Atacama Large Millimeter/submillimeter
  Array (ALMA), for which the highly excited rotational transitions of
  CO will be shifted outside the accessible (sub--)millimeter bands.
\end{abstract}


\section{Introduction}

Forbidden atomic fine structure transitions are important cooling
lines of the interstellar medium (ISM). They provide effective cooling
in cold regions where allowed atomic transitions can not be excited,
thus providing critical diagnostic tools to study the star--forming
ISM.  Perhaps the most important cooling line is the forbidden
$^2$P$_{3/2}\rightarrow^2$P$_{1/2}$ fine-structure line of ionized
carbon ([CII]) at 158 microns. Other main cooling lines are the oxygen
[OI] (63 microns) and [OIII] (at 52 and 88 microns) lines, as well as
the nitrogen [NII] lines at 122 and 205 microns (hereafter \nii).
Observations of a combination of these lines have proven to provide
key diagnostics regarding the physical properties of the atomic
interstellar medium in galaxies (e.g., probes for interstellar UV
radiation field, photon dominated regions [PDRs], temperature, gas
density and mass, e.g., Petuschowski \& Bennett 1993, van der Werf
1999). As the ionization potential of carbon is 11.3\,eV (hydrogen:
13.6\,eV), [CII] is a tracer for both the neutral atomic and ionized
medium, predominantly of PDRs.  The ionization potentials for oxygen
and nitrogen, on the other hand, are 13.6\,eV and 14.53\,eV,
respectively, implying that their forbidden ionized fine structure
lines trace the ionized medium only.  The \nii\ line, which is the
focus of this study, is of particular interest as it has a critical
density that is very close to that of [CII], thus potentially
providing complementary information on the origin of the [CII]
emission (e.g.  Oberst et al.\ 2006).

Given the poor atmospheric transmission at mid--/far--infrared
wavelengths most of of these lines in the local universe can only be
studied using airborne or space--based observatories. At sufficiently
high redshift, however, some of the lines are shifted to the
(sub--)millimeter bands that can be observed from the ground.
Especially at redshifts approaching the Epoch of Reionization (z$>$7)
these lines provide the only tools by which to study the properties of
the star--forming ISM using (sub--)millimeter facilities, as the
excited rotational lines of the standard tracer, CO, will be shifted
outside the observable (sub--)millimeter bands (Walter \& Carilli
2008).

Whereas the [CII] line has now been abundantly detected in the local
universe (e.g., Stacey et al.\ 1991, Malhotra et al.\ 1997, Luhman et
al.\ 1998, Madden et al.\ 1997), measurements of the \nii\ line are
scarce.\footnote{This is due to the fact that [CII] was accessible
  with the Infrared Space Observatory Long Wavelength Spectrometer
  (ISO/LWS, Clegg et al.\ 1996) whereas \nii\ was not.} The line was
first detected by {\em FIRAS} aboard {\em COBE} in the Milky Way
(Wright et al.\ 1991), where a global [CII]/\nii\ flux ratio of
10.4$\pm$1 was observed (first laboratory measurements were obtained
by Brown et al.\ 1994). The \nii\ line was later detected in the
Galactic HII region G333.6--0.2 (Colgan et al.\ 1993) and subsequently
in the Carina Nebula (Oberst et al. 2006); the first detection from
the ground (using SPIFI on AST/RO). Again a flux ratio around 10
(9.2$\pm$3.0) relative to the [CII] line was found.  The \nii\ line
was detected in only one extragalactic system, M\,82, using the
Kuiper Airborne Observatory, with a [CII]/\nii\ flux ratio of
20.0$\pm$3.4 (Petuchowski et al.\ 1994). At high redshift, the ratio
may be susceptible to N/C abundance variations (e.g., Matteucci \&
Padovani 1993). Given the potential importance of the \nii\ line in
studies of the ISM in the very early universe, multiple attempts were
performed to detect this line at high redshift (4C41.17 and
PC2047+0123: Ivison \& Harris 1996, the Cloverleaf: Benford 1999,
APM\,08279: Krips et al.  2007, see Table~1 for a summary).

Here we present a sensitive search for \nii\ emission in one of the
highest redshift quasars, SDSS J114816.64+525150.3 (hereafter:
J1148+5251) at z=6.42. It has a far--infrared (FIR) luminosity of
2.2$\times10^{13}$ L$_\odot$ (Bertoldi et al.\ 2003a, Beelen et al.\
2006) and hosts a large reservoir of molecular gas
(2\,$\times$\,10$^{10}$\,M$_\odot$), the prerequisite for star
formation, which has been detected through redshifted rotational
transition lines of CO (Walter et al.\ 2003, Bertoldi et al.\ 2003b,
Walter et al.\ 2004).  Besides BR\,1202--0725 (Iono et al.\ 2006), it
is the only system detected in [CII] at z$>$0 to date (Maiolino et
al.\ 2005, Walter et al.\ 2008) and is thus an obvious target to
search for emission from \nii. We use a concordance, flat $\Lambda$CDM
cosmology in this letter, with $H_0$=71\,km\,s$^{-1}$\,Mpc$^{-1}$,
$\Omega_{\rm M}$=0.27 and $\Omega_{\rm \Lambda}$=0.73 (Spergel et al.\
2007), and a resulting luminosity distance of 63.8\,Gpc at z=6.42
(Wright 2006).

\begin{figure}
\epsscale{1}
\plotone{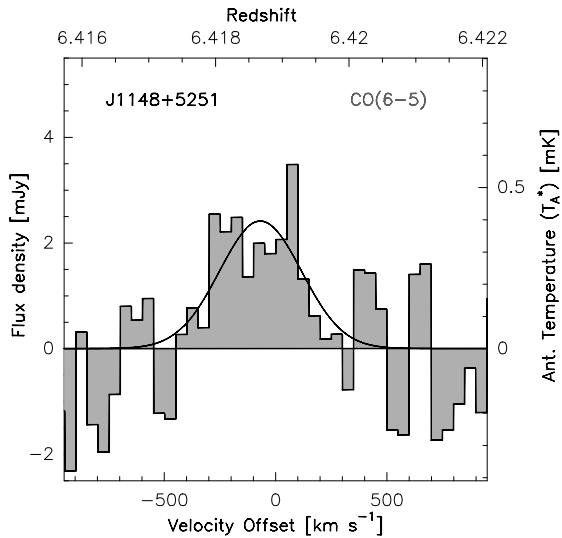}
\caption{The CO(6--5) line in J1148+5251 from the data that were 
taken in parallel to the \nii\ observations. Velocities are given
offset to the tuned redshift of z=6.4189. The best Gaussian fit is
overplotted as a solid curve (see Sec.~3.1 for fit parameters).
\label{f2}}
\end{figure}

\section{Observations}

J1148+5251 (z=6.4189, Bertoldi et al.\ 2003b) was observed with the
IRAM 30m telescope in March and April 2007 in good weather conditions.
We used the AB and CD receiver setups, with the AB receivers tuned to
CO(6--5) (3mm band, $\nu_{\rm obs}=93.20426$\,GHz, $\nu_{\rm rest}$=
691.473\,GHz) and the C/D receivers simultaneously tuned to \nii\ (1mm
band, $\nu_{\rm obs}=196.9472$\,GHz, $\nu_{\rm rest}$=1461.132\,GHz,
Brown et al. 1994). System temperatures were typically 130\,K at 3mm
and $\sim$210\,K at 1mm. Data were taken with a wobbler rate of 0.5\,Hz and
a wobbler throw of 50$"$ in azimuth. The pointing was checked
frequently and was found to be stable within 3$"$ during all runs.
Calibration was done every 12\,min with standard hot/cold load
absorbers, and we estimate fluxes to be accurate to within
$\sim$10--15\% in both bands.  We used the 512$\times$1\,MHz filter
banks for the 3mm receiver and the 256 $\times$ 4MHz filter banks for
the 1mm receivers.  As part of the data reduction, we dropped all
scans with distorted baselines, subtracted linear baselines from the
remaining spectra, and then rebinned to a velocity resolution of 50
km\,s$^{-1}$. The remaining useable `on--source' time is 11.0\,hr for
the CO(6--5) and 8.1\,hr for the \nii\ observations. The conversion
factors from K (T$_{\rm A}^*$ scale) to Jy at our observed frequencies
are 6.1\,Jy\,K$^{-1}$ at 3mm and 7.6\,Jy\,K$^{-1}$ at 1mm. We reached
an rms noise of 0.2\,mK (1.2~mJy) for CO(6--5) and of 0.17\,mK
(1.3~mJy) for the \nii\ observations.  In this letter we express line
luminosities both based on the source's flux density (L$_{\rm
  CO/[NII]}$ in units of L$_{\odot}$) and the source's brightness
temperature (L$'_{\rm CO/[NII]}$ in units of K\,km\,s$^{-1}$\,pc$^2$).
The reader is referred to the review by Solomon \& Vanden Bout (2005,
their equations 1 and 3) for a derivation of these quantities.

\begin{table*}
\normalsize
\begin{center}
\begin{tabular}{l|lllll}
  \hline
  source    &   S(\nii)$^A$      &  L$_{\rm [NII]205}$    &   L$_{\rm FIR}$     &   L$_{\rm [NII]205}$/L$_{\rm FIR}$  & reference \\
  &   Jy\,km\,s$^{-1}$ &  10$^8$\,L$_\odot$     &   10$^{12}$\,L$_\odot$&                                    &           \\
  \hline
  Cloverleaf&   $<$26.7          &  $<$4.6$^B$            &   5.6$^B$           &  $<$8.2\,10$^{-5}$                   & Benford 1999\\
  APM08279  &   $<$5.1           &  $<$4.6$^C$            &   4.6$^C$           &  $<$1.0\,10$^{-4}$                   & Krips et al.\ 2007\\
  4C41.17   &   $<$120           &  $<$447                &   15                &  $<$3.0\,10$^{-3}$                   & Ivison et al.\ 1996\\
  J1148+5251&   $<$0.48          &  $<$4.0               &   22                &  $<$2.0\,10$^{-5}$                   & this study\\
\end{tabular}
\caption{High--redshift [NII] limits from the literature.\\Table notes: $^A$: 3$\sigma$ upper limits for the following
  line--widths: Cloverleaf: 416\,km\,s$^{-1}$ (Barvainis et al.\ 1997,
  Wei{\ss} et al.\ 2003, 2005), APM\,08279: 480\,km\,s$^{-1}$ (Downes
  et al.\ 1999), 4C41.17: 1000\,km\,s$^{-1}$ (de Breuck et al.\ 2005),
  $^B$ corrected for a lensing magnification factor of $\mu$=11
  (Venturini \& Solomon 2003), $^C$ Assuming L$_{\rm FIR}$ of the
  `cold' dust component (Wei{\ss} et al.\ 2007a), corrected for a
  lensing magnification of $\mu$=4.2 (Riechers et al. 2009).}
\end{center}
\end{table*}

\section{Results}

\begin{figure}
\epsscale{1}
\plotone{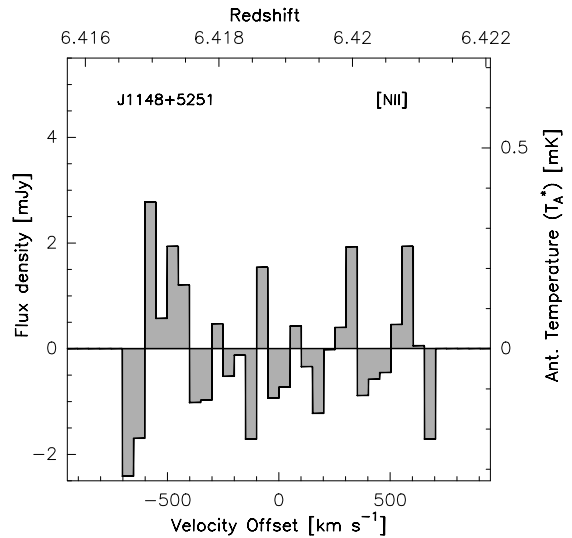}
\caption{Spectrum of the \nii\ observations towards J1148+5251 of the
  velocity and redshift range (lower and upper x--axis, respectively),
  expected for the \nii\ line. The line is not detected.
  \label{f3}}
\end{figure}

\subsection{CO(6--5) observations}

The CO(6--5) spectrum is shown in Fig.~1. The line is detected and
Gaussian fitting gives a redshift of z=6.4187$\pm$0.0002, a peak flux
density of 2.42\,mJy (0.4\,mK), a line width of
430$\pm$100\,km\,s$^{-1}$, leading to a flux integral of
1.12$\pm$0.3\,Jy\,km\,s$^{-1}$. Our peak flux is in good agreement
with the measurement by Bertoldi et al.\ (2003b) using the Plateau de
Bure interferometer (PdBI), however our line--width is larger
(430\,km\,s$^{-1}$ vs.\ 279\,km\,s$^{-1}$), leading to a flux integral
that is 50\% higher than the one quoted by Bertoldi et al. We
attribute the difference to the lower signal--to--noise ratio of our
observations. Adopting our measurements, we derive a CO(6--5) line
luminosity of L$'_{\rm
  CO}$=4.2$\pm$1.0$\times10^{10}$\,K\,km\,s$^{-1}$\,pc$^{2}$, or
L$_{\rm CO(6-5)}$=4.4$\pm$1.0$\times10^8$\,L$_{\odot}$.


\subsection{[NII] observations}

The \nii\ spectrum is shown in Fig.~2. No line emission is detected at
the sensitivity of our observations (1.3\,mJy in a 50\,km\,s$^{-1}$
channel). If we assume a linewidth of 300\,km\,s$^{-1}$ this results
in a 1$\sigma$ rms of 0.53\,mJy. The 3$\sigma$ upper limit for the
integrated flux thus is
3$\times$0.53\,mJy$\times$300\,km\,s$^{-1}$=0.48\,Jy\,km\,s$^{-1}$.
This translates to a 3$\sigma$ upper limit for the \nii\ line
luminosity of L$'_{\rm
  [NII]}\!<$4.0$\times10^{9}$\,K\,km\,s$^{-1}$\,pc$^{2}$, or L$_{\rm
  [NII]}\!<$4.0$\times10^8$\,L$_{\odot}$. The L$_{\rm [NII]}$/L$_{\rm
  FIR}$ flux ratio is thus $<\!2\!\times\!10^{-5}$ and the L$_{\rm
  [NII]}$/L$_{\rm CO(6-5)}$ ratio is $<\!0.9$.

\section{Summary and Discussion}

\subsection{Comparison to [CII] and Implications}

The ionization potential of carbon of 11.3\,eV is below that of
hydrogen (13.6\,eV), whereas the one for nitrogen is above that value
(14.53\,eV). The \nii\ and [CII] transitions have nearly identical,
low critical densities (\nii: 44\,cm$^{-3}$, [CII]: 46\,cm$^{-3}$,
assuming an electron temperature of 8000\,K, e.g. Oberst et al. 2006),
i.e.\ their line ratio is given by the N$^+$/C$^+$ abundance ratio in
the ionized medium. As pointed out by Oberst et al.\ (2006), this
ratio is insensitive to the hardness of the radiation field as the
energy levels of the next ionization states (N$^{++}$ and C$^{++}$)
are also similar (29.6 and 24.4 eV, respectively).

Our \nii\ observations are overplotted on the [CII] spectrum obtained
at the PdBI (Walter et al.\ 2009) in Fig.~3. The integrated flux of
the [CII] measurement is I$_{\rm
  [CII]}$=3.9$\pm$0.3\,Jy\,km\,s$^{-1}$, i.e. an order of magnitude
higher than our \nii\ upper limit. In terms of luminosities, the [CII]
luminosity is L$_{\rm [CII]}$=4.69$\pm$0.35$\times10^9$\,L$_\odot$, a
factor of 12 higher than the our upper limit for L$_{\rm [NII]}$.  Our
derived upper limit for the [CII]/\nii\ flux ratio thus is comparable
to the values found in the Milky Way (Wright et al.\ 1991) and M\,82
(Petuchowski et al.\ 1994) and sets tighter limits on the [CII]/\nii\
ratio than previous measurements at high redshifts (Table~1). Given
the non--detection of the \nii\ line in J1148+5251 and in other
sources at high redshift we can only speculate whether or not high--z
detections are within reach of currently operating observatories.

\subsection{Outlook}

Probing back into the Epoch of Reionization (z$>$7), forbidden atomic
fine structure lines will likely be the only means by which to
directly constrain the star--forming interstellar medium using
(sub--)millimeter interferometers. This is due to the fact that only
very high rotational transitions of CO can be observed at these
redshifts using (sub--)millimeter facilities. However, strong emission
from the highest (J$>6$) CO levels requires very highly excited gas
(either due to high kinetic temperatures, high densities, or both)
that is typically not present in large quantities in normal
starforming environments (e.g. Daddi et al.\ 2008).  Therefore high--J
CO lines are faint and difficult to detect in such systems.  Recent
studies of the CO excitation in high--redshift galaxies have shown
that the CO excitation ladder (`CO SEDs') peak at around J$\sim$6 for
QSOs and J$\sim$5 for submillimeter galaxies (Wei{\ss} et al.\ 2005,
2007a, 2007b, Riechers et al.\ 2006).

\begin{figure}
\epsscale{1}
\vspace*{6mm}
\plotone{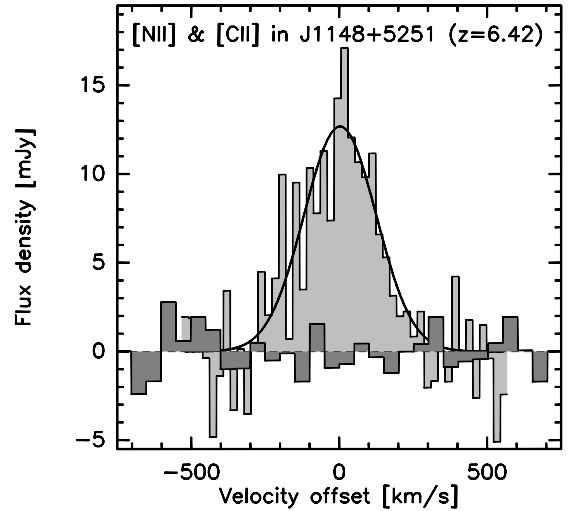}

\caption{Comparison of the \nii\ non--detection towards J1148+5251 and
  the [CII] emission (continuum subtracted) in the same source (Walter
  et al.\ 2009).
\label{f3}}
\end{figure}

\begin{figure*}
\epsscale{1.0}
\plotone{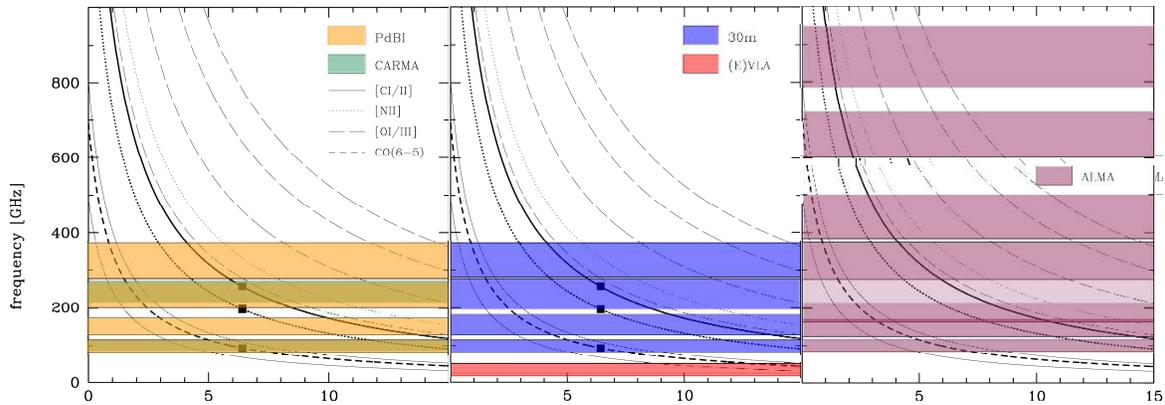}
\caption{Frequency coverage of redshifted fine structure lines for
  some current (and future) key telescopes as a function of redshift.
  The following lines are plotted (in order of increasing frequency,
  given in GHz]): CI$\nu$492.2, CO(6--5)$\nu$691.5, CI$\nu$809.3,
  [NII]$\nu$1461.1 (=\nii), [CII]$\nu$1900.5, [OI]$\nu$2060,
  [NII]$\nu$2460, [OIII]$\nu$3393, [OI]$\nu$4745, [OIII]$\nu$5786).
  The {\em left} panel shows the coverage for the PdBI and CARMA, the
  {\em middle} panel the respective coverages for the EVLA and the
  IRAM 30\,m telescope and the {\em right} panel for ALMA (bands 2--9,
  band 5 is not fully funded and thus appears in lighter color). The
  three black points in the left and middle panels highlight the
  positions of the transitions in J1148+5251 discussed in this letter
  (CO(6--5), \nii\ and [CII]).
\label{f4}}
\end{figure*}

Figure 4 shows which atomic fine strucure lines can be observed in a
given frequency band of some (present and future) telescopes as a
function of redshift.  The CO(6--5) transition (which lies close to
the peak of the highly excited QSOs) is also shown for comparison. It
is obvious from this figure that at z$>$7, the atomic fine structure
lines will be critical probes to constrain the properties of the
interstellar medium using (sub--)millimeter instruments. Redshifted
mm--wavelength molecular line observations have so far only probed the
molecular medium of the target sources. Fine structure lines of
various species, and in particular their luminosity ratios, will also
yield information on the neutral and ionized gas that is more dilute.

Even though we have not detected the \nii\ line in our observations,
we have reached a limit that is consistent with what is found in our
Galaxy and in the starburst galaxy M\,82. This may indicate that we
are close to being able to detect atomic fine structure lines other
than [CII] at high redshift --- however it is not clear if detections
can be obtained given currently operating facilities. In the ALMA era,
these lines will provide the fundamental tools needed to constrain the
physical properties of the star--forming ISM in the earliest epoch of
galaxy formation. These observations will then complement fine
structure line measurements in the nearby universe using future
airborne and space missions (e.g., {\em SAFIRE} on {\em SOFIA} and,
e.g., {\em PACS} onboard {\em HERSCHEL}).

\vspace{5mm}

The authors thank Roberto Maiolino, Alberto Bolatto and Helmut
Wiesemeyer for useful comments on the manuscript. FW and DR
acknowledge the hospitality of the Aspen Center for Physics, where
parts of this manuscript were written.  DR acknowledges support from
NASA through Hubble Fellowship grant HST-HF-01212.01-A awarded by the
Space Telescope Science Institute, which is operated by the
Association of Universities for Research in Astronomy, Inc., for NASA,
under contract NAS 5-26555. CC acknowledges support from the
Max-Planck-Gesellschaft and the Alexander von Humboldt-Stiftung
through the Max-Planck-Forschungspreis 2005.

\end{document}